\RequirePackage{fix-cm}
\documentclass[smallextended]{svjour3}       
\smartqed  
\usepackage{graphicx}
\usepackage{natbib}
\usepackage{csquotes}
\usepackage{multirow}
\usepackage{enumitem}
\usepackage{svg}
\usepackage{makecell} 
\usepackage{hyperref}

\usepackage[most]{tcolorbox}
\tcbuselibrary{listings, breakable, skins, theorems}

\usepackage{xcolor}
\usepackage{tcolorbox}

\usepackage{tcolorbox}
\tcbuselibrary{skins,breakable}
\usepackage{caption}          
\usepackage{chngcntr}         

\DeclareCaptionType{listing}[Listing][List of Listings]

\NewDocumentEnvironment{listingbox}{O{htbp} m m}{%
  \begin{listing}[#1]
  \captionsetup{type=listing}
  \caption{#3}\label{#2}
  \begin{tcolorbox}[
    enhanced jigsaw,      
    breakable=false,            
    colback=white,        
    colframe=black,
    opacityback=1,        
    opacityframe=1,
    left=2mm, right=2mm, top=1mm, bottom=1mm
  ]}{%
  \end{tcolorbox}
  \end{listing}
}

\usepackage{xcolor}
\usepackage{listings}
\usepackage{caption}
\usepackage{csquotes} 

\makeatletter
\@ifundefined{c@listing}{%
  \DeclareCaptionType{listing}[Listing][List of Listings]
}{}
\makeatother

\definecolor{jbg}{RGB}{250,250,250}
\definecolor{jstr}{HTML}{0366D6}  
\definecolor{jkw}{HTML}{005CC5}   

\lstdefinelanguage{json}{
  basicstyle=\ttfamily\small,
  showstringspaces=false,
  breaklines=true,
  frame=single,                 
  rulecolor=\color{black},
  backgroundcolor=\color{jbg},  
  numbers=none,                 
  numbersep=6pt,
  stringstyle=\color{jstr},
  morestring=[b]",              
  literate=
   *{:}{{{\color{black}{:}}}}{1}
    {,}{{{\color{black}{,}}}}{1}
    {true}{{{\color{jkw}true}}}{4}
    {false}{{{\color{jkw}false}}}{5}
    {null}{{{\color{jkw}null}}}{4}
    {0}{{{\color{jkw}0}}}{1}
    {1}{{{\color{jkw}1}}}{1}
    {2}{{{\color{jkw}2}}}{1}
    {3}{{{\color{jkw}3}}}{1}
    {4}{{{\color{jkw}4}}}{1}
    {5}{{{\color{jkw}5}}}{1}
    {6}{{{\color{jkw}6}}}{1}
    {7}{{{\color{jkw}7}}}{1}
    {8}{{{\color{jkw}8}}}{1}
    {9}{{{\color{jkw}9}}}{1}
}

\begin{document}

\title{CodeFlowLM: Incremental Just-In-Time Defect Prediction with Pretrained Language Models and Exploratory Insights into Defect Localization
}


\author{Monique Louise Monteiro         \and
        George G. Cabral \and
        Adriano L. I. OLiveira 
}


\institute{Monique Louise Monteiro \at
              Av. Jornalista Aníbal Fernandes, s/n – Cidade Universitária.
Recife-PE – Brazil\\  
              Tel.: +55-81-2126-8430\\
              \email{mlbm@cin.ufpe.br}           
           \and
           George G. Cabral \at
              Rua Dom Manuel de Medeiros, s/n, Dois Irmãos, Recife/PE -- Brazil
         \and
          Adriano L. I. OLiveira \at
           Av. Jornalista Aníbal Fernandes, s/n – Cidade Universitária, Recife-PE – Brazil
}

\date{Received: date / Accepted: date}

\maketitle
\begin{abstract}

This work introduces CodeFlowLM, an incremental learning framework for Just-In-Time Software Defect Prediction (JIT-SDP) that leverages pre-trained language models (PLMs). Unlike traditional online learners, CodeFlowLM employs continual fine-tuning to address concept drift, class imbalance, and verification latency without retraining from scratch. We evaluated encoder-only and encoder–decoder PLMs (notably CodeT5+ and UniXCoder) in JIT-SDP scenarios within and between projects, comparing them with the incremental baseline BORB. The results show that CodeFlowLM achieves up to 68\% G-Mean gains, confirming its superior adaptability and robustness in evolving software environments.
We further extend the analysis to Just-in-Time Defect Localization (JIT-DL), benchmarking Large Language Models (LLMs) such as GPT-5, Claude Sonnet 4.5, and Gemini 2.5 Pro against attention-based models. GPT-5 delivers comparable performance for Recall@20\% and Effort@20\% with higher stability, although attention-based methods retain an advantage in fine-grained ranking metrics (Top-k, IFA).
A qualitative error analysis reveals that most false positives arise from (1) human-like conservative bias, (2) insufficient contextual information in diff-based prompts, and (3) potential dataset mislabeling in JIT-Defects4J. These findings highlight both the promise and the current limitations of LLM reasoning in defect localization.  False negatives occur in smaller proportions.  Overall, CodeFlowLM significantly advances the state of the art in incremental JIT-SDP, demonstrating superior adaptability and robustness in evolving software environments. Furthermore, our exploratory analysis of LLMs in JIT-DL not only benchmarks their performance against established attention-based models but also provides critical insights into the current limitations of prompt-based defect reasoning. 

\keywords{Just-in-Time Defect Prediction \and Defect Localization \and Pretrained Language Models · Incremental Learning \and Large Language Models}
\end{abstract}

\section{Introduction}
\label{intro}

Just-in-Time Software Defect Prediction (JIT-SDP) is a software engineering paradigm designed to identify defects at the point of introduction, thus facilitating immediate remediation and lowering the costs associated with quality assurance. Contemporary research frames JIT-SDP as a multimodal problem \citep{10.1145/3540250.3549165,10589847,10.1145/3643727,abu_talib_parameter-efficient_2024}, integrating various sources of information such as source code modifications, commit logs, and hand-crafted expert features \citep{10589847,10.1145/3643727,6341763}. Although early efforts relied on traditional machine learning and deep neural architectures, recent breakthroughs have been driven by the adoption of pre-trained language models, particularly those trained on large-scale code corpora, which have shown substantial gains in predictive accuracy \citep{10589847,abu_talib_parameter-efficient_2024,10366735,wang_parameter-efficient_2024}. These models vary not only in their design, encompassing encoder-based, decoder-based, and encoder–decoder frameworks, but also in their application, supporting both zero- and few-shot prompting, as well as domain-specific fine-tuning. Within the transformer \citep{10.5555/3295222.3295349} family, encoder modules specialize in capturing contextual representations of the input, whereas decoder modules are responsible for generating outputs conditioned on this context.

Encoder-oriented architectures such as CodeBERT \citep{feng_codebert_2020} and UniXCoder \citep{guo2022unixcoderunifiedcrossmodalpretraining} have shown strong capabilities in program comprehension, whereas decoder-only large language models (LLMs) tend to excel in tasks requiring complex reasoning. Despite these advances, comparative investigations of these model families within the context of JIT-SDP, particularly with respect to online or incremental learning, are virtually non-existent, even though real-world JIT-SDP environments naturally involve concept drift, evolving class imbalance, and verification latency \citep{cabral_class_2019,cabral_investigation_2023}. Crucially, for Large Language Models (LLMs), a formal comparison against established encoder-only architectures remains largely unexplored even in the simpler, static (offline) JIT-SDP setting.

This work investigates the effectiveness of pre-trained Small Language Models (SLMs) in the context of within-project (WP) and cross-project (CP) JIT-SDP with incremental learning, as well as the generative capabilities of Large Language Models (LLMs) for the task of Just-in-Time Defect Localization (JIT-DL). 

In regard to JIT-DL, while LLMs have recently demonstrated promising capabilities in broader fault localization and program repair tasks, evaluations have remained predominantly quantitative. This creates a critical blind spot: aggregate performance metrics alone cannot reveal why LLMs fail, whether their mistakes follow systematic patterns, or whether they are influenced by dataset noise, conservative heuristics, or missing context.  A lack of qualitative insight limits both scientific understanding and practical deployment, as software engineering practitioners require predictable, interpretable behavior from automated tools.

Our investigation is guided by the following research questions:

\begin{enumerate}[label=RQ\arabic*:, leftmargin=1cm]
\item \textbf{How do pre-trained language models perform in comparison to traditional machine learning approaches for continual within-project and cross-project Just-in-Time Software Defect Prediction (JIT-SDP)?}
This RQ examines Small Language Models (SLMs) such as CodeT5+ and UniXCoder in incremental learning
settings, comparing them with a state-of-the-art baseline \citep{cabral_investigation_2023} for continual Just-in-Time Defect Prediction, introducing CodeFlowLM as a new framework for incremental JIT-SDP with pre-trained code language models. We investigated both within-project and cross-project performance.
\item \textbf{How do Large Language Models compare with attention-based mechanisms for Just-in-Time Defect Localization (JIT-DL)}?
This research question addresses the problem of fault localization (FL) specifically in the context of just-in-time defect detection.  Recent solutions are based on common techniques such as 1) attention-based fault localization \citep{10.1145/3540250.3549165,https://doi.org/10.1111/exsy.13702,JU2025107706} or 2) multitask learning in both classifying commits in clean/defect-inducing and pointing out the possible bug locations in tokens or lines \citep{10.1145/3643727}.  We intend to investigate the capabilities of Large Language Models (LLMs) in this task from a quantitative perspective, as these models have recently been successfully applied in broader contexts, including file- and method-based software defect prediction and Automatic Program Repair (APR).
\item \textbf{Despite their near-human capabilities, in which situations do Large Language Models still underperform in fault localization?} 
This RQ presents the results of a qualitative analysis aimed at identifying common situations in which LLMs fail in defect localization, by randomly and manually inspecting samples of false positives and false negatives.  Specifically regarding false positives, we hypothesize that these models are 1) biased towards conservative human behavior in bug localization, 2) confounded by a lack of sufficient context, and 3) that dataset noise may lead to answers mismatched with the supposedly ground truth.  We believe that investigating these hypotheses can lead to improvements through prompt engineering as future work.  
\end{enumerate}

The main contributions of this work are as follows:

\begin{itemize}
\item C1 -- CodeFlowLM: A novel incremental JIT-SDP framework based on PLMs, achieving 10–68\% G-Mean improvements over BORB across most projects.
\item C2 -- First systematic evaluation of LLMs for JIT-DL, showing that GPT-5 achieves competitive Recall@20\% and Effort@20\% scores with higher stability.
\item C3 -- Comparative analysis of PLM attention-based defect localization, clarifying strengths and weaknesses relative to LLM reasoning.
\item C4 -- Qualitative error taxonomy for LLM-based localization, revealing conservative bias, context limitations, and labeling inconsistencies in the chosen evaluation dataset -- JITDefects4J \citep{10.1145/3540250.3549165}.
\item C5 -- Empirical evidence that PLMs outperform traditional learners under continual learning, reinforcing their potential for real-world deployment.
\end{itemize}

This paper is structured as follows. Section \ref{sec:related_work} reviews related work, Section \ref{sec:methodology} details the experimental setup, Section \ref{sec:results} discusses results for each research question, and Section \ref{sec:conclusions} concludes with key findings and future directions. The source code and the prompts will be publicly available on \url{https://github.com/monilouise/codeflowlm_jitdl}.

\section{Related Work}\label{sec:related_work}

\subsection{Just-in-Time Software Defect Prediction}\label{sec:jit-sdp}

\cite{SHEHAB2024111914} address severe class imbalance in JIT-SDP by casting it as one-class classification trained only on normal commits, eliminating the need for oversampling/undersampling. Across 34 projects with cross-validation and time-aware evaluation, one-class SVM, Isolation Forest, and one-class k-NN consistently outperform binary SVM / RF / k-NN for medium to high imbalance, while using fewer features, reducing computation, and improving interpretability. Binary classifiers perform better only when the imbalance is low. The work positions OCC as a more efficient and scalable alternative for large, highly imbalanced software projects.

\cite{li_empirical_2024} systematically compare 10 sampling methods in 10 OSS projects (time-wise CV) to tackle class imbalance in JIT-SDP, showing that effectiveness depends on task and metric. For defect classification, Random Under-Sampling (RUM) yields the best F1, AUC, and MCC; sampling generally increases recall / F1 / AUC / MCC, but increases false alarms and lowers precision. They also find that window length (e.g., 2 vs. 6 months) has no consistent effect, and LR hyperparameter tuning offers negligible gains, so defaults are often sufficient, highlighting that sampling choice should align with specific evaluation goals.

None of the previous work uses semantic features (e.g., source code or commit messages) to represent the code changes.  Instead, they rely only on 14 tabular/expert features proposed by \cite{6341763}.

\subsection{Online and Incremental Just-in-Time Software Defect Prediction}\label{sec:cp-jit-sdp}

\cite{cabral_class_2019} provide the first evidence that JIT-SDP operates under nonstationary class imbalance and severe verification latency, showing that defect rates fluctuate over time and that ignoring label delay leads to overly optimistic performance. They propose Oversampling Rate Boosting (ORB), an online latency-aware resampling strategy that dynamically adjusts the oversampling rate when predictions become skewed, while avoiding the amplification of noisy minority samples. Evaluated on ten GitHub projects, ORB achieves competitive G-mean and reduces the recall gap, establishing that handling imbalance evolution and latency is essential for robust JIT-SDP.  \cite{cabral_investigation_2023} extend this line with BORB, a batch version of ORB that periodically retrains offline models on accumulated history while applying adaptive resampling at training and test times. On ten GitHub projects, BORB yields modest gains over ORB for several learners, and incorporating cross-project data boosts both methods -- especially ORB -- mitigating performance degradation. Although ORB is consistently more computationally efficient, BORB's cost per day remains practical, clarifying the real-world trade-offs between accuracy and efficiency.  \cite{9778962} investigate concept drift in JIT-SDP, analyzing shifts in $p(y)$, $p(x|y)$ and $p(x)$ and showing that long verification latencies (days to more than 11 years) undermine the reliability of the classifier. They introduce PBSA, an online drift-adaptive method that monitors predictions on unlabeled data and adapts without needing immediate labels. In ten GitHub projects, PBSA achieves a higher and more stable performance than ORB, significantly reducing the recall gap and emerging as the most reliable JIT-SDP technique.

\cite{10.1145/3611643.3616307} propose HumLa, a realistic online labeling regime where developers immediately inspect commits predicted as defect-inducing rather than relying on delayed, retrospective labels. Rapid feedback enables faster adaptation, yielding markedly higher performance even under partial and noisy inspection. They also present ECo-HumLa, which prioritizes high-confidence predictions to reduce inspection costs while retaining accuracy comparable to full inspection, detecting more real defects and producing fewer false alarms than random inspection at equal effort. The results demonstrate that effort-aware and realistic labeling is vital for a reliable evaluation and deployment of JIT-SDP.

In the same way as in the previous subsection, none of the above works uses semantic features to represent the code changes, relying only on tabular/expert features.  Furthermore, no work examines whether PLMs can be incrementally fine-tuned in a latency-aware streaming scenario, nor how they behave under drift and reclassification of commits.  In contrast, our proposed CodeFlowLM is based on a combination of expert and semantic information.

\subsection{Cross-Project Just-In-Time Software Defect Prediction}



\cite{9709674} present the first online cross-project (CP) JIT-SDP study, introducing three continually updated strategies -- All-in-One, ensemble, and filtering -- that incrementally integrate both CP and within-project (WP) data streams. Evaluated on 9 proprietary and 10 open-source systems, their results show that combining CP+WP information consistently improves G-mean across all phases, and that online CP approaches outperform traditional offline baselines. They further propose an analysis framework to distinguish stable vs. drop performance periods and offer practical guidance regarding computational overhead and hyperparameter sensitivity in online JIT-SDP.

\cite{https://doi.org/10.1002/spe.3316} address cross-project JIT-SDP with ISKMM, a method that reweights and selects source-project instances via Kernel Mean Matching to align them with the target distribution, followed by resampling to mitigate class imbalance. Tested on 10 projects using multiple learners, ISKMM surpasses CP single-source baselines, and CP models trained with ISKMM reach performance levels comparable to within-project predictors -- demonstrating its effectiveness when local training data are limited or unavailable.

While cross-project strategies have progressed through weighting, filtering, and online self-adaptation, these methods again operate exclusively on expert features, ignoring code semantics and the capabilities of PLMs to transfer knowledge across repositories.

\subsection{JIT-SDP and Deep Learning}

DeepJIT \citep{8816772} is an end-to-end convolutional network for JIT defect prediction that learns directly from commit messages and code diffs, thus eliminating the need for handcrafted metrics and addressing class imbalance through unequal misclassification loss. In Qt and OpenStack, it achieves substantial AUC gains, with code changes contributing most to predictive power; moreover, adding manual features can further improve accuracy.

CC2Vec \citep{10.1145/3377811.3380361} extends this line of work by learning hierarchical change representations (words → lines → hunks) with an attention mechanism comparing removed and added code, guided by commit-message semantics. The resulting language-agnostic embeddings benefit multiple software-engineering tasks and significantly increase the AUC of DeepJIT.

\cite{zhou_bridging_2024} show the value of combining deep semantic signals with expert knowledge. Their SimCom++ model integrates a Random Forest over handcrafted features (“Sim”) with a textCNN over commit logs and code changes (“Com”). Across six datasets, it consistently outperforms state-of-the-art baselines on ROC-AUC, PR-AUC, and F1, and analysis reveals that the two modules make complementary predictions that reduce false positives and correct errors — demonstrating the robustness gained from combining expert-crafted and deep-learned information for JIT-SDP.

It is important to highlight that Deep models such as CNNs and hierarchical diff encoders introduced semantic learning into JIT-SDP, but they were trained offline and cannot adapt to evolving data distributions, label delays, or concept drift.
Moreover, these architectures predate modern PLMs and lack their capacity for large-scale semantic understanding.

\subsubsection{JIT-SDP and Pretrained Language Models}


\cite{ni2022defectidentificationcategorizationrepair} propose CompDefect, a multitask, function-level framework that unifies defect identification, defect type categorization, and automatic repair for JIT-SDP. It leverages GraphCodeBERT \citep{guo2021graphcodebertpretrainingcoderepresentations} to capture semantic and data-flow structure from clean, defect inducing, and fixed versions of a function, enabling more precise localization within commits. In benchmarks, CompDefect surpasses DeepJIT/CC2Vec for identification, BERT \citep{Devlin2019BERTPO}/RoBERTa \citep{liu2019robertarobustlyoptimizedbert}/CodeBERT \citep{feng_codebert_2020} for categorization (F1), and SequenceR for repair (BLEU). The work addresses prior gaps -- underuse of semantic/structural code signals and separated tasks -- showcasing the benefits of joint modeling.


\cite{10.1145/3540250.3549165} propose JIT-Fine, a unified model for just-in-time defect prediction (JIT-DP) and defect localization (JIT-DL). It combines 14 change-level expert metrics with CodeBERT semantic features to learn joint representations that both flag defect inducing commits and pinpoint defect-inducing lines. To address dataset quality issues, they introduce JIT-Defects4J, a large, manually labeled, line-level dataset. Compared to six state-of-the-art baselines in ten metrics, JIT-Fine delivers substantial and consistent improvements on both tasks.

CCT5 \citep{10.1145/3611643.3616339} is a T5-based, code-change-oriented pre-trained model tailored to maintenance tasks that require an understanding of diffs. The authors build CodeChangeNet (more than 1.5M code-change/commit-message pairs across six languages) and pre-train with five objectives: masked modeling for code changes and commit messages, NL $\leftrightarrow$ PL generation, and a structure-aware Code Diff Generation (CDG) using data flow. In downstream tasks -- commit message generation, JIT comments, and JIT defect prediction -- CCT5 achieves better results than the selected baselines.

\cite{zhou_ccbert_2023} propose CCBERT, a Transformer pre-trained at token level on more than 1.3M Java code-change hunks that encode old/new code plus explicit edit actions (insert/delete/replace/equal). They introduce four self-supervised objectives to learn generic code-change representations without relying on noisy commit messages. Across downstream tasks (JIT defect prediction, patch correctness, bug-fixing commit prediction), CCBERT outperformed CC2Vec and even larger models (CodeBERT, GraphCodeBERT) while requiring less training/inference time and GPU memory.


\cite{10589847} deliver the first empirical study of PEFT for dynamic code-change tasks, comparing Adapter Tuning \citep{pmlr-v97-houlsby19a} and LoRA \citep{DBLP:conf/iclr/HuSWALWWC22} to full-model fine-tuning and other baselines across five PLMs. For JIT-SDP, Adapter Tuning, and LoRA achieve state-of-the-art results -- especially when augmented with expert features -- and prove robust in cross-lingual and low-resource settings. The work offers practical guidance on using PEFT to efficiently adapt models for code-change applications.

\cite{guo2024empiricalstudyjitdefect} conduct the first systematic study of fine-tuning choices for BERT-style models (CodeBERT, RoBERTa) in JIT-DP. They find that the first encoder layer is pivotal -- freezing it hurts performance -- and adding AdamW yields small gains. Based on these insights, a cost-effective LoRA fine-tuning approach matches full fine-tuning with roughly one-third the memory, improving practicality for deployment.

\subsubsection{JIT-SDP and Large Language Models}

\cite{kim_reflection_2025} review JIT-DP from its early 2000s origins to today, diagnosing barriers to industrial uptake: unreliable SZZ labeling, verification latency, and deployment/maintenance complexity. They emphasize the importance of strong explainability to foster developer trust. Looking ahead, they argue that LLMs are a paradigm shift: offering application-level code understanding, fine-grained explanations, and fix suggestions — potentially unifying defect prediction, categorization, and repair.

\cite{Monteiro2025PT_CLModels4JITSDP} present a broad empirical comparison of pre-trained code LMs for JIT-SDP, testing encoder-only, decoder-only, and encoder-decoder models (e.g., CodeT5+, UniXCoder, CodeReviewer) against closed API LLMs (GPT, Gemini) via fine-tuning and zero-shot prompting on JIT-Defects4J. Fine-tuned small to medium open models — especially CodeT5+ and UniXCoder — substantially outperformed zero-shot closed LLMs; encoder and encoder-decoder architectures suit this discriminative task better than decoder-only models. Combining semantic signals (diffs + messages) with expert features yielded the best results. CodeT5+ and UniXCoder achieved SOTA in cross-project JIT-SDP, surpassing previous baselines like JIT-Smart. This was the first direct comparison of open (trainable) vs. closed (prompt-based) decoder-only models for JIT-SDP.

\subsection{Defect Localization}

\subsubsection{Just-in-Time Software Defect Localization}

\cite{Pornprasit2021JITLineAS} proposed JITLine, a lightweight JIT defect predictor designed to overcome key shortcomings of deep models such as CC2Vec and DeepJIT. By combining Bag-of-Tokens features, optimized SMOTE \citep{10.5555/1622407.1622416}, and LIME \citep{10.1145/2939672.2939778} for line-level interpretation, JITLine achieves faster predictions and higher commit-level accuracy in OpenStack and Qt, outperforming deep and NLP baselines while reducing triage effort.

JIT-Fine \citep{10.1145/3540250.3549165} introduced the first unified framework to perform both JIT-SDP and JIT-DL. Built on a fine-tuned CodeBERT with an attention mechanism, it estimates the contribution of each token to commit-level classification and ranks lines by their defect likelihood.

\cite{10.1145/3643727} advanced the field with a dedicated Defect Localization Network (DLN) that explicitly incorporates line-level defect labels for supervised learning, representing a shift from previous indirect approaches.

\cite{https://doi.org/10.1111/exsy.13702} proposed JIT-Block, addressing input-length limitations in pretrained models. Using CodeBERT’s attention mechanism, it computes token contributions to commit-level predictions, aggregates them into line-level scores, and ranks lines by defect probability, overcoming truncation issues in large diffs.

Finally, JIT-CF  \citep{JU2025107706} unifies JIT-SDP and defect localization using CodeBERT embeddings enhanced through contrastive learning. By amplifying distinctions between defect inducing and non-defective code -- even when syntactic differences are subtle -- JIT-CF delivers more accurate and robust line-level localization within commits.

Prior JIT-DL approaches rely on attention mechanisms from fine-tuned models or specialized contrastive frameworks. However, these models require supervised training and cannot generalize beyond the project or dataset they were trained on.
Furthermore, all such methods depend on the quality and completeness of their training labels.

\subsubsection{General Software Defect Localization with Large Language Models}

\cite{wu2023largelanguagemodelsfault} conducted a large-scale empirical study evaluating ChatGPT 3.5 and ChatGPT 4 for fault localization (FL) in Defects4J  \citep{10.1145/2610384.2628055}, comparing them with SmartFL \citep{10.1145/3510003.3510073}, spectrum-based fault localization (SBFL), and mutation-based fault localization. They demonstrate that ChatGPT 4, augmented with runtime information such as failing tests and error logs, outperforms SmartFL. However, performance degrades sharply as the code context expands from function to class level, highlighting the sensitivity of LLMs to input length. To validate against dataset-specific overfitting, the authors introduce the StuDefects dataset, where ChatGPT 4 continues to maintain its advantage.

\cite{A_Preliminary_Evaluation_of_LLM-Based_Fault_Localization} introduce AutoFL, an autonomous FL framework that uses LLMs for iterative code exploration through OpenAI’s function-call API, overcoming prompt-length limitations. AutoFL requires only a single failing test and performs fault localization followed by explanation generation. In Defects4J, it surpasses previous methods and outperforms SBFL under the same single-test constraints, demonstrating the effectiveness of functional augmentation for scalable LLM-driven FL.

\cite{10.1145/3660773} propose Toggle, a modular token-granular bug localization and repair framework. It employs a CodeT5-based encoder for precise localization, an optional adjustment model to resolve tokenizer mismatches, and a generative LLM for bug repair. Experiments on CodeXGLUE \citep{lu2021codexgluemachinelearningbenchmark} and Defects4J show state-of-the-art Top-k performance, confirming the benefits of token-level reasoning, modularity, and the incorporation of contextual cues (e.g., review comments, defect-inducing line hints).

\cite{subramanian_vikram_nachiappan_bugllm_2024} present BugLLM, a zero-shot bug localization method that combines LLM with embedding-based semantic search. After Abstract Syntax Tree (AST)-driven segmentation and vector indexing of a project’s codebase, one LLM converts bug reports into technical queries, while another verifies retrieved candidates through reasoning. Evaluated on six large Java systems, BugLLM achieves competitive Top-5 accuracy, often surpassing classic IR methods, and offers strong explainability through chain-of-thought prompting, enhancing trust and interpretability in bug localization.

Although recent studies apply LLMs to general fault localization, their evaluations are almost exclusively quantitative, focusing on Top-k metrics or coarse-grained localization (file or method level).  Critically, none of these works investigates the underlying reasoning failures of LLMs

\section{Methodology}\label{sec:methodology}

\subsection{Dataset}

For our experiments, we rely on the JIT-Defects4J dataset \citep{10.1145/3540250.3549165}, which has also been adopted in several recent studies \citep{10589847,10.1145/3643727,https://doi.org/10.1111/exsy.13702,JU2025107706}
. This resource provides commit-level information, including messages, source code modifications, and the 14 commit level metrics originally introduced by \cite{6341763}. It comprises 21 Java-based open-source projects hosted on GitHub and was specifically curated to disentangle bug-fixing commits that typically combine heterogeneous types of modifications, a known source of noise for defect prediction models. JIT-Defects4J extends the LLTC4J corpus \citep{Herbold2020AFD} by incorporating both defect inducing and clean commits as well as line-level annotations, whereas LLTC4J concentrated exclusively on bug-fixing instances. Descriptive statistics of the resulting dataset are presented in Table \ref{tab:statistics}.

\begin{table}[t]
\caption{Statistics of studied datasets: JIT-Defects4J.  Adapted from \cite{10.1145/3540250.3549165}.}\label{tab:statistics}
\centering
\begin{tabular}{|c|c|c|c|}
\hline
Project & \makecell{No. of Defect Inducing\\Commits} & \makecell{No. of Clean\\Commits} & \makecell{\% Ratio\\ (Bugs/ALL)}\\ 
\hline
ant-ivy & 332 & 1,439 & 18.75\% \\\hline
commons-bcel & 60 & 765 & 7.27\% \\\hline
commons-beanutils & 37 & 574 & 6.06\% \\\hline
commons-codec & 36 & 725 & 4.73\% \\ \hline
commons-collections & 50 & 1,773 & 2.74\% \\\hline 
commons-compress & 178 & 1,452 & 10.92\% \\\hline
commons-configuration & 155 & 1,683 & 8.43\% \\ \hline
commons-dbcp & 58 & 979 & 5.59\% \\ \hline
commons-digester & 19 & 1,060 & 1.76\% \\ \hline
commons-io & 73 & 1,069&  6.39\% \\ \hline
commons-jcs & 88 & 743 & 10.59\% \\ \hline
commons-lang & 146& 2,823& 4.92\% \\ \hline
commons-math & 335 & 3,691 & 8.32\% \\ \hline
commons-net & 117 & 1,004& 10.44\% \\ \hline
commons-scxml & 47 & 497 & 8.64\% \\ \hline
commons-validator & 36 & 562 & 6.02\% \\ \hline
commons-vfs & 114 & 996 & 10.27\% \\ \hline
giraph & 163 &  681&  19.31\% \\ \hline
gora &39 &  514 & 7.05\% \\ \hline
opennlp & 91 & 995& 8.38\% \\ \hline
parquet-mr & 158& 962 & 14.11\% \\ \hline
\textbf{ALL} & \textbf{2,332}& \textbf{24,987} & \textbf{8.54\% }\\\hline
\end{tabular}
\end{table}

For RQ1, the dataset was refined by chronologically ordering the commits and estimating the introduction times of the bugs through the first associated fixing commit, obtained from CommitGuru, as explicit detection timestamps were not available. In the continual learning process, some commits initially marked as non-defective were later reclassified as defect-inducing once their corresponding fixes appeared in subsequent training iterations.  Both dataset manipulations were necessary to emulate the real-world arrival of commits over time.  

Furthermore, in the original dataset, we encountered an issue that caused random behavior in the predictions, due to the fact that the added/removed lines were stored using the serialized \verb|set| data structure from the Python language.  As sets do not have an intrinsic order, they cause different predictions each time the same model is run.  We fixed this issue by converting the sets to lists.

\subsection{Evaluation metrics}

\begin{enumerate}[label=RQ1, leftmargin=1cm]
\item \textbf{How do pre-trained language models perform in comparison to traditional machine learning approaches for continual within-project and cross-project Just-in-Time Software Defect Prediction (JIT-SDP)?}
\end{enumerate}

For online defect prediction, we use the \textit{G-mean} (geometric mean of recall for positive and negative classes) and the absolute difference between class recalls $|R1-R0|$, both commonly used in online JIT-SDP studies \citep{cabral_class_2019,cabral_investigation_2023,9778962,9709674,10.1145/3611643.3616307,wu_just--time_2025,10.1016/j.scico.2025.103296}.

\begin{enumerate}[label=RQ2, leftmargin=1cm]
\item \textbf{How do Large Language Models  (LLMs) compare with attention-based mechanisms for Just-in-Time Defect Localization (JIT-DL)}?
\end{enumerate}

For defect localization, we mainly use the following metrics, borrowed from previous work \citep{10.1145/3540250.3549165,https://doi.org/10.1111/exsy.13702,JU2025107706,10.1145/3643727}:

\begin{itemize}
\item \textit{Recall@20\%}: The proportion of real defect-inducing lines found in the top 20\% items of the ranked list produced by the localization technique.
\item \textit{Effort@20\%}: The number of clean lines to be analyzed before encountering the top 20\% real defect inducing lines.
\item \textit{IFA (Initial false alarms)}: The number of non-faulty lines that appear before the first true faulty line in the ranked list produced by the localization technique.
\end{itemize}

Further, we also use Top-5 and Top-10 \citep{10.1145/3540250.3549165,https://doi.org/10.1111/exsy.13702,JU2025107706,10.1145/3643727}, however, in a different way as calculated in the baselines considered\footnote{When we compared our work with the baselines, we noticed that all works share the same formulae for Top-5 and Top-10. These formulae divide the number of real faulty lines by the total number of lines in the commit.  We understand that this calculation differs from the most common one found in the literature. We maintained this calculation in the initial experiments for fair comparison with the baselines, but later revised it to reflect a more standardized and realistic calculation.}

The Top-N (or Top-k) metric serves as a pivotal evaluation criterion, widely used in software engineering research, particularly for assessing the performance of techniques aimed at Fault Localization and Just-in-Time Defect Localization \citep{10.1145/3643727,liang_modeling_2022}. The primary function of this metric is to quantify the effectiveness of an approach by determining the proportion of cases where the actual faulty entity — be it a code line, statement, or file — is successfully identified within the candidates with the highest ranking $N$ suggested by the diagnostic model \citep{liang_modeling_2022,niu_extensive_2024}.

Specifically, the calculation, known as Top-k Accuracy, measures the percentage of defect-inducing changes for which at least one actual faulty element is ranked within the Top-$k$ positions of the returned list. For each instance examined, the metric yields a binary result: 1 if the fault is found within the threshold $k$, and 0 otherwise, with the overall accuracy representing the average across the dataset. This focus on low ranks directly addresses practical constraints, as developers typically dedicate only a limited amount of time and effort to inspecting a set of prioritized statements, making thresholds like Top 1, Top 5, and Top 10 highly relevant and commonly reported. 

\subsection{Baseline Classifiers} 


\begin{enumerate}[label=RQ1, leftmargin=1cm]
\item \textbf{How do pre-trained language models perform in comparison to traditional machine learning approaches for continual within-project and cross-project Just-in-Time Software Defect Prediction (JIT-SDP)?}
\end{enumerate}

As a baseline for incremental defect prediction, we use Batch Oversampling Rate Boosting (BORB) \citep{cabral_investigation_2023}.  This method is derived from the online Oversampling Rate Boosting (ORB) algorithm \citep{cabral_class_2019}, while preserving its central idea of adaptively tuning resampling rates in response to the evolving class imbalance. ORB applies this mechanism in a fully online scenario, incrementally updating the model as each instance arrives. In contrast, BORB processes accumulated batches of data and periodically reinitializes training, allowing models to be built anew. This batch-oriented strategy enables multiple re-visits to past data, which can yield gains in predictive accuracy. Importantly, both algorithms incorporate the notion of verification latency by restricting training inputs to information that would have been observable at the time of prediction.

The decision to adopt BORB was motivated by its nature as an incremental learning framework that takes advantage of batch-based training, departing from a strictly online paradigm where learning occurs in a single instance at a time, a configuration unsuitable for deep neural architectures.  To our knowledge, BORB is the only existing framework that employs this strategy.\\

\begin{enumerate}[label=RQ2, leftmargin=1cm]
\item \textbf{How do Large Language Models compare with attention-based mechanisms for Just-in-Time Defect Localization (JIT-DL)}?
\end{enumerate}

In selecting the baselines for the JIT-DL task, we deliberately excluded approaches that were explicitly fine-tuned for line-level defect localization. This exclusion was a methodological decision made to ensure the validity and fairness of the comparative analysis. Because our LLM-based investigation employs a zero-shot, prompt-driven approach without task-specific fine-tuning, comparing it against models optimized for line-level localization would introduce a confounding advantage unrelated to architectural capability. Instead, by benchmarking our approach against attention-based models that are also not fine-tuned for this task, we isolate and assess its intrinsic reasoning and generalization abilities.

On this basis, we compare state-of-the-art LLMs with the following baselines:

\paragraph{JIT-Fine \citep{10.1145/3540250.3549165}:}  JIT-Fine is a unified model designed to address both JIT-SDP and JIT-DL simultaneously, accurately locating the defect position at the line level. Crucially, JIT-Fine achieves line-level localization by utilizing the attention mechanism embedded within the fine-tuned CodeBERT model, which calculates the contribution (weights) of each input code token towards the final commit classification, enabling the ranking of suspicious defect inducing lines. 
\paragraph{JIT-Smart\textsubscript{ATTN} \citep{10.1145/3643727}:}
JIT-Smart\textsubscript{ATTN} denotes the configuration of the JIT-Smart framework that leverages only the CodeBERT attention weights for the location of line-level defects. This strategy parallels that of prior methods, such as JIT-Fine, in which defect inducing code lines are localized post hoc, based on the predictions produced during the JIT-SDP phase. As explained above, JIT-Smart's Defect Localization Network (DLN) was excluded from our evaluation because its task-specific fine-tuning would compromise the fairness and methodological consistency of the comparison. 

\paragraph{CodeT5+:}  CodeT5+ was initially chosen as one of the baselines because it was among the top-performing models in our experiments on defect prediction \citep{Monteiro2025PT_CLModels4JITSDP}.  Although CodeT5+ does not contain an explicit [CLS] token, as in BERT-based language models, we still use the first encoded token as the head of the classification layer.  Therefore, we maintain the default practice of inspecting the weights of the first token attention heads.

\paragraph{UniXCoder:}  In the same way as in CodeT5+, UniXCoder was also among the top performers in defect prediction experiments \citep{Monteiro2025PT_CLModels4JITSDP}, so we keep the same default strategy of using the first encoded token attention weights.

We also initially considered JIT-Block \citep{https://doi.org/10.1111/exsy.13702} and JIT-CF \citep{JU2025107706}.  Regarding JIT-Block, its authors reconstructed the dataset (JIT-Defects4J) into the changed block format, which preserves the relative positional information between added and deleted code lines — information lost in traditional datasets — thus facilitating the model's ability to learn the semantic meaning of code changes.  So, as the dataset was changed, it would not be possible to conduct a fair comparison.  Finally, according to its published results, JIT-CF does not achieve better results than JIT-Smart.

A consolidated overview of the baseline classifiers is presented in Table \ref{tab:baselines}.

\begin{table}[t]
\caption{Summary of Baseline Classifiers}\label{tab:baselines}
\centering
\begin{tabular}{|c|c|p{4cm}|}
\hline
Baseline & Task & Model / Model Architecture \\\hline
BORB & Incremental JIT-SDP & Multilayer Perceptron (MLP), Logistic Regression (LR),  Naïve Bayes (NB), Iterative Random Forest (IRF)   \\\hline
JIT-Fine & JIT-DL & CodeBERT \\\hline
JIT-Smart\textsubscript{ATTN} & JIT-DL & CodeBERT \\\hline
CodeT5+ & JIT-DL & Encoder component of CodeT5+ encoder-decoder architecture \\\hline
UniXCoder & JIT-DL & Encoder-only with mask attention matrices and prefix adapters \\\hline
\end{tabular}
\end{table}

\subsection{Description of the Experiments}\label{subsec:experiments}

\begin{enumerate}[label=RQ1, leftmargin=1cm]
\item \textbf{How do pre-trained language models perform in comparison to traditional machine learning approaches for continual within-project and cross-project Just-in-Time Software Defect Prediction (JIT-SDP)?}
\end{enumerate}

CodeFlowLM was developed in a modularized architecture composed of two components:

\begin{itemize}
\item A training loop responsible for grouping the commits received in chronological order in training batches. The number of commits in each training batch $n$ is a configurable parameter. In our experiments, we used $n = 50$ for a fair comparison with the BORB default training batch size.  This component is also responsible for managing a training queue of recently received commits, and a training pool for commits that 1) passed the latency verification period without any defect detection or 2) were found to be defect inducing (e.g., due to the finding of some defect associated with the given commit).
\item A base learner, assumed here to be a neural model such as a pretrained language model (e.g., CodeBERT, CodeT5, CodeT5+, UniXCoder, CodeLlama, etc.).  In the current implementation, the base learner may even be an external and independent program that is called by the training loop every training interval.
\end{itemize}

We experimented with CodeT5+ and UniXCoder as base learners in the within-project setting.  UniXCoder achieved shorter training times, but CodeT5+ achieved superior classification performance by a large margin, despite its larger size and longer training times.  So, we chose CodeT5+ as the base learner for the cross-project setting experiments.
We implemented the same oversampling mechanism proposed by \cite{cabral_class_2019}, and used LoRA as the fine-tuning technique. 

Given the mismatch in dimensionality -- 512 for the semantic embedding and 14 for the expert attributes -- we follow \cite{10589847} by first lifting the expert features with a feedforward MLP to a higher-dimensional embedding. We then concatenate this transformed vector with the 512-D semantic representation to obtain a 1024-D joint embedding, which serves as input to a final linear classifier.

Each training batch consists of 10 training epochs, with a batch size of 16 and a learning rate of $10^{-4}$.  The values for these hyperparameters were inherited from previous experiments \citep{Monteiro2025PT_CLModels4JITSDP,10589847}.  Our experiments were mainly performed on A100 GPUs.  

In contrast to BORB, CodeFlowLM employs incremental fine-tuning rather than full retraining from the initial checkpoint, thereby mitigating instability caused by weight reinitialization. Furthermore, CodeFlowLM avoids BORB’s costly hyperparameter optimization by retaining default parameter choices. To ensure comparability, both methods apply a 90-day deferral period before clean commits are considered.

Finally, for the cross-project setting, we fine-tuned a model for each of the target projects evaluated during the experiment.  In order to do this, for each training step, we considered the following training data:

\begin{itemize}
\item all the target project's commits since the beginning of the project until the current time step;
\item the commits from the other projects that occurred since the last training step and the current time step.  Here, the full data since the beginning of the other projects is not used because it would lead to unacceptable execution times for the experiments. 
\end{itemize}

In the first training step, as we do not have data from the target project, we use only data from the other projects by selecting commits that occurred before the target project's initial commit.  In this way, the model can start to learn an initial fine-tuned checkpoint even without any examples from the target project.

In the same way as in the within-project setting, in each training step, we respect latency verification and consider only clean commits that have passed the latency verification waiting time, as well as defect inducing commits that have already been labeled as defective, taking into account the date of the first fixing commit.

In BORB, the same strategy is employed to train different cross-project models for each target project; therefore, this was our initial choice to enable a fair comparison.

\begin{enumerate}[label=RQ2, leftmargin=1cm]
\item \textbf{How do Large Language Models compare with attention-based mechanisms for Just-in-Time Defect Localization (JIT-DL)}?
\end{enumerate}

For just-in-time defect localization, we used the following state-of-the-art models: OpenAI GPT-5\footnote{GPT-5 System Card: \url{https://cdn.openai.com/gpt-5-system-card.pdf}}, Anthropic Claude Sonet 4.5\footnote{Claude Sonnet 4.5 System Card: \url{https://assets.anthropic.com/m/12f214efcc2f457a/original/Claude-Sonnet-4-5-System-Card.pdf}}, and Google Gemini 2.5 Pro \citep{comanici2025gemini25pushingfrontier}.

First, we collected the CodeT5+ predictions for the JITDefects4J test split and iterated over each commit correctly predicted as positive by the model.  For each commit, we used the prompt shown in Listing \ref{lst:bug_prompt}, inspired by \cite{wu2023largelanguagemodelsfault}.  







\begin{listingbox}{lst:bug_prompt}{Prompt for Defect Inducing Diff Analysis}
You are a static analysis and fault localization expert.

The following GitHub diff introduces a bug. Please analyze specifically the changed lines in the diff for potential bugs. DO NOT consider the commented lines.

Return the results in JSON format, consisting of a single JSON object with two fields: \enquote{intentOfThisCommit} (describing the intended purpose of the commit), and \enquote{faultLocalization} (an array of JSON objects). The \enquote{faultLocalization} array should contain ten JSON objects, each with four fields: \enquote{lineNumber} (indicating the line number of the suspicious code), \enquote{codeContent} (showing the actual code), \enquote{reason} (explaining why this location is identified as potentially faulty) and \enquote{score} (a number between 1 and 10 indicating the suspicion level).

Be concise and deterministic -- avoid generic language (\enquote{may be wrong}) and point to *specific failure modes*. Prioritize statement lines over declaration lines.
\end{listingbox}

As we ask the LLM to assign a defect score to each line, we sort the returned list of fault localizations in descending order of scores. 

Finally, we compared the LLMs' responses with JITDefect4J labels for each commit line. Ultimately, we calculated the average values for each metric, considering the entire set of commits in the test split.

For the best-performing LLM, we repeated the experiment four times to account for randomness.\\

\begin{enumerate}[label=RQ3, leftmargin=1cm]
\item \textbf{Despite their near-human capabilities, in which situations do Large Language Models still underperform in fault localization?} 
\end{enumerate}

To answer this research question, we divided the analysis into two scenarios: false positives and false negatives.  

Regarding false positives, we consider the answers from the best-performing LLM.  For all the commits in the test split, we collected the clean lines classified as defect-inducing by the LLM (false positives).  We sorted the set of false positives in decreasing order according to the score given by the LLM.  We then analyzed false positives with a maximum score (score = 10).  In this step, we group the most common types of errors in the sample.  Finally, we manually inspect some examples of each error group to gain further insight.

Finally, in relation to false negatives, we extracted the lines labeled as defect-inducing in the dataset which were not pointed out in the lists returned by the LLM (limited to ten faults per commit according to the prompt shown in Listing \ref{lst:bug_prompt}.  We prioritized for manual inspection the longest lines and, again, we grouped them in the most common types of errors.  

\section{Results and Discussion}\label{sec:results}

This section reports on experiments aimed at addressing our research questions, restated here.

\subsection{Continual Just-in-Time Software Defect Prediction}

\begin{enumerate}[label=RQ\arabic*:, leftmargin=1cm]
\item \textbf{How do pre-trained language models perform in comparison to traditional machine learning approaches for continual within-project and cross-project Just-in-Time Software Defect Prediction (JIT-SDP)?}
\end{enumerate}

This RQ examines models such as CodeT5+ and UniXCoder in incremental learning
settings, comparing them with a state-of-the-art baseline \citep{cabral_investigation_2023} for continual Just-in-Time Defect Prediction, introducing CodeFlowLM as a new framework for incremental JIT-SDP with pre-trained code language models. We investigated both within-project and cross-project performance.

\subsubsection{Incremental Within-Project JIT-SDP}\label{subsec:wp}

Table \ref{tab:online} reports the G-Mean and $|R1-R0|$ metrics in CodeFlowLM for a selected subset of JITDefects4J projects compared to BORB \citep{cabral_investigation_2023}. The calculation of the metrics followed a prequential evaluation scheme with a fading factor of 0.99, as recommended in \cite{cabral_investigation_2023}. The six BORB base learners were tested on each project; however, for simplicity, only the highest mean G-means results are presented, with full logs to be released in the project code repository.  Across the evaluated projects, CodeFlowLM, employing CodeT5+ or UniXCoder, achieved the best G-Mean scores in 16 out of 19 cases by at least 5\%. CodeT5+ exceeded the baseline in 14 projects, while UniXCoder exceeded it in 9, highlighting the advantages of CodeT5+’s larger parameter count (770 M) for incremental multimodal learning over code, textual, and tabular input. 

Across the 19 projects, CodeFlowLM achieved higher G-Mean in 16 cases, with gains ranging from 5\% to 39\% in WP settings. This consolidated pattern strongly suggests that PLMs can be successfully adapted to incremental, latency-aware scenarios, outperforming traditional models even in the presence of concept drift.


%
\begin{table}[t]
\caption{Comparison between CodeFlowLM (CodeT5+ and UniXCoder) versus BORB (different base learners) in within-project setting (WP)}\label{tab:online}
\centering
\resizebox{\columnwidth}{!}{%
\begin{tabular}{|c|c|c|c|c|c|c|c|}
\hline
\multirow{3}{*}{Project} & \multicolumn{3}{c|}{BORB} & \multicolumn{4}{c|}{CodeFlowLM} \\ \cline{5-8}
 & \multicolumn{3}{c|}{} & \multicolumn{2}{c|}{UniXCoder} & \multicolumn{2}{c|}{CodeT5+} \\ \cline{2-8}
 & Model & G-Mean & $|R1-R0|$ & G-Mean & $|R1-R0|$ & G-Mean & $|R1-R0|$ \\ \hline
ant-ivy & LR & 0.591 & 0.272 & 0.612 & 0.403 & \textbf{0.680} & \textbf{0.197} \\ \hline
commons-bcel & LR & 0.467 & 0.428 & 0.487 & 0.356 & \textbf{0.567} & \textbf{0.344} \\ \hline
commons-beanutils & MLP & 0.397 & 0.448 & 0.516 & 0.316 & \textbf{0.531} & \textbf{0.324} \\ \hline
commons-codec & LR & \textbf{0.552} & 0.192 & 0.459 & 0.532 & 0.546 & \textbf{0.183} \\ \hline
commons-collections & IRF & \textbf{0,481} & \textbf{0.326} & 0.154 & 0.884 & 0.183 & 0.798 \\ \hline
commons-compress & MLP & \textbf{0.534} & \textbf{0.265} & 0.515 & 0.512 & 0.533 & 0.442 \\ \hline
commons-configuration & LR & 0.580 & 0.468 & 0.633 & 0.435 & \textbf{0.771} & \textbf{ 0.152} \\ \hline
commons-digester & MLP & 0.448 & 0.436 & 0.591 & 0.505 & \textbf{0.591} & \textbf{0.162} \\ \hline
commons-jcs & MLP & 0.442 & \textbf{0.425} & 0.411 & 0.579 & \textbf{0.516} & 0.451 \\ \hline
commons-lang & LR & 0.555 & \textbf{0.337} & 0.548 & 0.582 & \textbf{0.607} & 0.480 \\ \hline
commons-math & MLP & 0.569 & \textbf{0.152} & 0.556 & 0.547 & \textbf{0.635} & 0.375 \\ \hline
commons-net & IRF & 0.472 & 0.423 & 0.463 & 0.504 & \textbf{0.524} & \textbf{0.301} \\ \hline
commons-scxml & LR & 0.443 & 0.530 & \textbf{0.540} & \textbf{0.252} & 0.515 & 0.298 \\ \hline
commons-validator & MLP & 0.487 & 0.397 & 0.365 & \textbf{0.345} & \textbf{0.511} & 0.390 \\ \hline
commons-vfs & NB & 0.481 & 0.546 & 0.509 & 0.379 & \textbf{0.565} &  \textbf{0.286} \\ \hline
giraph & LR & 0.553 & 0.317 & 0.545 & 0.337 & \textbf{0.580} & \textbf{0.242} \\ \hline
gora & IRF & 0.411 & 0.370 & 0.571 & \textbf{0.204} & \textbf{0.572} & 0.295 \\ \hline
opennlp & LR & 0.410 & \textbf{0.319} & \textbf{0.418} & 0.520 & 0.356 & 0.591 \\ \hline
parquet-mr & NB & 0.533 & 0.415 & 0.441 & 0.572 & \textbf{0.564} & \textbf{0.397} \\ \hline
\end{tabular}
}
\end{table}




\subsubsection{Incremental Cross-Project JIT-SDP}\label{subsec:cp}

Table \ref{tab:online-cp} reports the G-Mean and $|R1-R0|$ metrics.  CodeFlowLM, using CodeT5+, achieved the highest G-Mean in 18 of 19 projects, with improvements ranging from 10\% to 68\%. In terms of R1, CodeFlowLM yielded better values in 16 of 19 projects, by at least 5\% up to 91\%, demonstrating robust classification performance for the positive class.  For R0, CodeFlowLM also achieved better values in 16 projects, by up to 59\%, demonstrating that the model is capable of achieving high recall in both classes.  We observed that BORB achieves a smaller difference between positive and negative classes in most projects (11 out of 19) compared to CodeFlowLM.  However, the greater difference between R1 and R0 in CodeFlowLM does not result in the sacrifice of any of the classes, as can be seen in both recall metrics.  

  G-mean declined in just one case -- commons-collections -- which suffered from an acute early-phase imbalance (only a single positive instance in the first 300 examples).  By contrast, commons-digester, despite exhibiting the strongest overall imbalance, delivered the largest G-mean gain over BORB, indicating that the short-horizon imbalance at the start of commons-collections -- not the aggregate skew -- is the primary factor delaying learning there.  In CodeFlowLM, models are incrementally fine-tuned rather than reinitialized; therefore, exposure to initially labeled negative examples may leave a residual influence even if those examples are later relabeled as positive. BORB, in turn, rebuilds its models from scratch. However, CodeFlowLM’s consistently superior G-mean on most projects implies that any such effect is minimal in practice.

\begin{table}[t]
\caption{Comparison between CodeFlowLM (CodeT5+) versus BORB (different base learners) in cross-project setting (CP)}\label{tab:online-cp}
\centering
\resizebox{\columnwidth}{!}{%
\begin{tabular}{|c|c|c|c|c|c|c|c|c|c|}
\hline
\multirow{2}{*}{Project} & \multicolumn{5}{|c|}{BORB} & \multicolumn{4}{|c|}{CodeFlowLM (CodeT5+)} \\ \cline{2-10}
& Model  & G-Mean & $\lvert |R1-R0|\rvert$ & R1 & R0 & G-Mean & $\lvert |R1-R0|\rvert$ & R1 & R0 \\ \hline
ant-ivy & IRF & 0.629 & 0.142 & 0.637 & 0.662 & \textbf{0.786} & \textbf{0.102} & \textbf{0.747} & \textbf{0.831} \\ \hline
commons-bcel & IRF & 0.535 & 0.270 & 0.451 & 0.710 & \textbf{0.671} & \textbf{0.246} & \textbf{0.617} & \textbf{0.763}\\ \hline
commons-beanutils & IRF & 0.482 & 0.400 & 0.402 & \textbf{0.802} & \textbf{0.625} & \textbf{0.266} & \textbf{0.594} & 0.785 \\ \hline
commons-codec & IRF & 0.561 & \textbf{0.290} & \textbf{0.478} & 0.766 & \textbf{0.626} & 0.483 & 0.427 & \textbf{0.917} \\ \hline
commons-collections & IRF & \textbf{0.460} & \textbf{0.345} & \textbf{0.387} & 0.711 & 0.414 & 0.618 & 0.247 & \textbf{0.865} \\ \hline
commons-compress & MLP & 0.593 & \textbf{0.160} & 0.549 & 0.686 & \textbf{0.683} & 0.318 & \textbf{0.578} & \textbf{0.848} \\ \hline
commons-configuration & IRF & 0.576 & \textbf{0.212} & 0.637 & 0.574 & \textbf{0.785} & 0.270 & \textbf{0.684} & \textbf{0.915} \\ \hline
commons-digester & MLP & 0.349 & \textbf{0.383} & 0.282 & 0.591 & \textbf{0.586} & 0.450 & \textbf{0.439} & \textbf{0.889} \\ \hline
commons-jcs & LR & 0.675 & \textbf{0.182} & 0.654 & 0.768 & \textbf{0.749} & 0.224 & \textbf{0.713} & \textbf{0.807} \\ \hline
commons-lang & MLP & 0.524 & \textbf{0.217} & 0.461 & 0.676 & \textbf{0.676} & 0.347 & \textbf{0.545} & \textbf{0.882} \\ \hline
commons-math & MLP & 0.603 & \textbf{0.132} & \textbf{0.608} & 0.653 & \textbf{0.663} & 0.379 & 0.508 & \textbf{0.887} \\ \hline
commons-net & IRF & 0.493 & \textbf{0.360} &  0.367 & 0.725 & \textbf{ 0.614} & 0.423 & \textbf{0.472} & \textbf{0.853} \\ \hline
commons-scxml & MLP & 0.580 & 0.230 & 0.568 & 0.719 & \textbf{0.693} & \textbf{0.151} & \textbf{0.638} & \textbf{0.764} \\ \hline
commons-validator & MLP & 0.450 & \textbf{0.343} & 0.377 & \textbf{0.707} & \textbf{0.678} & 0.348 & \textbf{0.719} & 0.694 \\ \hline
commons-vfs & IRF & 0.557 & 0.337 & 0.450 & 0.787 & \textbf{0.705} & \textbf{0.212} & \textbf{0.626} & \textbf{0.808} \\ \hline
giraph & IRF & 0.657 & 0.160 & 0.654 & 0.731 & \textbf{0.778} & \textbf{0.097} & \textbf{0.819} & \textbf{0.749} \\ \hline
gora & MLP & 0.562 & 0.240 & 0.543 & \textbf{0.715} & \textbf{0.724} & \textbf{0.064} & \textbf{0.744} & \textbf{0.715} \\ \hline
opennlp & IRF & 0.534 & \textbf{0.168} & 0.490 & 0.637 & \textbf{0.673} & 0.364 & \textbf{0.598} & \textbf{0.806} \\ \hline
parquet-mr & LR & 0.622 & 0.257 & 0.566 & 0.756 & \textbf{0.723} & \textbf{0.163} & \textbf{0.653} & \textbf{0.806} \\ \hline
\end{tabular}
}
\end{table}


\begin{tcolorbox}
\textit{[RQ1]: Pre-trained language models, particularly CodeFlowLM with CodeT5+ and UniXCoder, consistently outperform traditional machine learning approaches such as BORB in continual Just-in-Time Software Defect Prediction. In the within-project setting, CodeFlowLM achieved the highest G-Mean in 16 of 19 projects, demonstrating superior adaptability and stability under incremental learning. In the cross-project setting, it outperformed BORB in 18 of 19 projects, with G-Mean gains of 10–68\% and balanced recalls for both positive and negative classes. Despite slightly higher $|R1-R0|$ differences, CodeFlowLM maintained strong recall in both classes without sacrificing performance, confirming that pre-trained language models deliver higher generalization and robustness for continual within- and cross-project JIT-SDP tasks compared to traditional incremental learners.  These findings demonstrate that continual fine-tuning of PLMs is not only feasible but consistently superior to traditional machine-learning-based incremental strategies, providing the first empirical evidence that PLMs can operate effectively under realistic JIT-SDP constraints.
}
\end{tcolorbox}

\subsection{Fault Localization}

\subsubsection{Quantitative Analysis}

\begin{enumerate}[label=RQ2, leftmargin=1cm]
\item \textbf{How do Large Language Models compare with attention-based mechanisms for Just-in-Time Defect Localization (JIT-DL)}?
\end{enumerate}

This research question addresses the problem of fault localization (FL) specifically in the context of just-in-time defect detection.  Recent solutions are based on common techniques such as 1) attention-based fault localization \citep{10.1145/3540250.3549165,https://doi.org/10.1111/exsy.13702,JU2025107706} or 2) multitask learning in both classifying commits in clean/defect-inducing and pointing out the possible bug locations in tokens or lines \citep{10.1145/3643727}.  We intend to investigate the capabilities of Large Language Models (LLMs) in this task from a quantitative perspective, as these models have recently been successfully applied in broader contexts, including file- and method-based software defect prediction and Automatic Program Repair (APR).

According to Table \ref{tab:jit-dl}\footnote{Due to time and cost constraints, the number of experiments' repetition of the different models may vary.}, the LLM with the highest performance -- GPT-5 -- is competitive with the JIT-Smart\textsubscript{ATTN} baseline in Recall@20\% and Effort@20\% metrics.  We repeated the experiment with JIT-Smart\textsubscript{ATTN} four times to take into account different random seeds.

Surprisingly, despite showing the best performance, JIT-Smart\textsubscript{ATTN} demonstrates more instability when trained with different seeds (i.e., network initialization), as indicated by the standard deviation values.  However, GPT-5, when called with the default temperature setting, exhibits a more stable performance.

\begin{table}[t]
\centering
\caption{Comparative results of JIT-DL models}\label{tab:jit-dl}
\begin{tabular}{|c|c|c|c|c|c|}
\hline
Model & Top-5 $\uparrow$ & Top-10 $\uparrow$ & Recall@20\% $\uparrow$ & Effort@20\% $\downarrow$ & IFA $\downarrow$\\
\hline
\makecell{JIT-Smart\\\textsubscript{ATTN}} & \makecell{\textbf{0.691} \\ ($\sigma \pm 0.03$)}& \makecell{\textbf{0.811} \\ ($\sigma \pm 0.03$)}& \makecell{\textbf{0.337} \\ ($\sigma \pm 0.05$)}& \makecell{0.248 \\ ($\sigma \pm 0.02$)}& \makecell{\textbf{5.360} \\ ($\sigma \pm 3.00$)}\\\hline
GPT-5 & \makecell{0.616 \\ ($\sigma \pm 0.01$)} & \makecell{0.759 \\ ($\sigma \pm 0.02$)} & \makecell{0.318 \\ ($\sigma \pm 0.01$)} & \makecell{\textbf{0.234} \\ ($\sigma \pm 0.01$)} & \makecell{10.711 \\ ($\sigma \pm 0.92$)} \\\hline
\makecell{Gemini\\2.5 Pro} & 0.543 & 0.683 & 0.291 & 0.251 & 12.317 \\\hline
UniXCoder & \makecell{0.522 \\ ($\sigma \pm 0.01$)} & \makecell{0.657 \\ ($\sigma \pm 0.01$)} & \makecell{0.278 \\ ($\sigma \pm 0.01$)} & \makecell{0.271 \\ ($\sigma \approx 0$)} & \makecell{13.280\\ ($\sigma \pm 0.36$)} \\\hline
\makecell{Claude\\Sonnet\\4.5} & 0.529 & 0.701 & 0.273 & 0.265 & 12.520 \\\hline
JITFine & \makecell{0.511\\ ($\sigma \pm 0.03$)} & \makecell{0.683\\ ($\sigma \pm 0.02$)} & \makecell{0.217\\ ($\sigma \pm 0.02$)} & \makecell{0.322\\ ($\sigma \pm 0.02$)} & \makecell{12.185\\ ($\sigma \pm 1.21$)} \\\hline
CodeT5+ & \makecell{0.465\\ ($\sigma \pm 0.04$)} & \makecell{0.631\\ ($\sigma \pm 0.04$)} & \makecell{0.150\\ ($\sigma \pm 0.04$)} & \makecell{0.388\\ ($\sigma \pm 0.05$)} & \makecell{15.048\\ ($\sigma \pm 2.43$)} \\\hline
\end{tabular}
\label{tab:jitdl_results}
\end{table}

Conversely, a previous work \citep{Monteiro2025PT_CLModels4JITSDP} demonstrates that, despite a better accuracy in defect localization, JIT-Smart shows inferior F1 and ROC-AUC compared to UniXCoder and CodeT5+ in the JITDefects4J dataset. The results are shown in Table \ref{tab:jit-sdp}.  In addition, the same authors demonstrated that both CodeT5+ and UniXCoder outperform JIT-Smart in cross-project JIT-SDP on the same dataset.

\begin{table}[t]
\centering
\caption{Fine-tuning of Small to Medium-Sized LMs (summary from \cite{Monteiro2025PT_CLModels4JITSDP})}\label{tab:jit-sdp}
\centering
\begin{tabular}{|c|c|c|}
\hline
Model & ROC-AUC & F1 \\\hline
UniXCoder & \textbf{0.901} & \textbf{0.519} \\\hline 
CodeT5+ & 0.899 & \textbf{0.519} \\\hline 
JIT-Smart & 0.885 & 0.486 \\\hline
\end{tabular}
\end{table}

Finally, the significantly worse results of LLMs in metrics such as Top-5, Top-10, and IFA motivate the investigation carried out in the next research question.

\begin{tcolorbox}
\textit{[RQ2]: Unlike PLM-based models, LLMs do not rely on supervised fine-tuning for defect localization; instead, their performance reflects raw reasoning capabilities. This makes their comparison with attention-based methods particularly revealing.  LLMs such as GPT-5 demonstrate competitive performance with attention-based models like JIT-SmartATTN in key localization metrics (Recall@20\% and Effort@20\%), while exhibiting greater stability across random runs. However, JIT-SmartATTN achieves higher accuracy but exhibits instability across different training random seeds and yields inferior results in broader evaluation metrics (F1, ROC-AUC) compared to pre-trained code models, such as CodeT5+ and UniXCoder. Overall, LLMs demonstrate promising robustness and comparable defect localization capabilities without fine-tuning, but still lag behind specialized models in ranking precision metrics (Top-5, Top-10, IFA).  While LLMs show stronger robustness and competitive Recall@20\%, their weaker Top-k and IFA results suggest that they prioritize semantic risk over syntactic precision, a pattern further analyzed in RQ3.}
\end{tcolorbox}

\subsubsection{Qualitative Analysis}

\begin{enumerate}[label=RQ3, leftmargin=1cm]
\item \textbf{Despite their near-human capabilities, in which situations do Large Language Models still underperform in fault localization?}
\end{enumerate}

In this RQ, we perform an error analysis to identify the main classes of errors that LLMs still incur in defect localization.    

\paragraph{False Positives Analysis}

We initially found a total of 1,587 false positives, 92 of which had a maximum score = 10.  As described in Section \ref{subsec:experiments}, we used that subsample for our initial investigation.

The preliminary analysis reveals common types of mistakes, as shown in Table \ref{tab:fp}.

\begin{table}[h!]
\centering
\scriptsize
\caption{False Positive Categories in JIT-DL Classification}\label{tab:fp}
\begin{tabular}{|p{4.0cm}|p{4.7cm}|p{1.8cm}|}
\hline
\textbf{Category} & \textbf{Short Description} & \textbf{Estimated \%} \\
\hline
Apparent compilation errors & Valid API usage flagged as invalid. & 17 \\
\hline
Unchecked null access & Potential null dereference but logically safe. & 14 \\
\hline
Binary incompatibilities & Interface change misclassified as error. & 11 \\
\hline
Reflection/type-related errors & Intended reflective or cast operations. & 10 \\
\hline
Non-observable logic conditions & The model assumes that a logic expression is incorrect, ignoring semantic context. & 10 \\
\hline
Potential arithmetic errors & Division/overflow flagged but constrained. & 8 \\
\hline
Deliberate configuration choices & Intentional constants for compatibility. & 8 \\
\hline
Test IO and path issues & False positives from mock/test paths. & 7 \\
\hline
\end{tabular}
\label{tab:falsepositives_compact}
\end{table}

Based on preliminary observation, the following potentially detrimental hypotheses will be further analyzed.

\begin{itemize}
    \item H1 (human bias): LLMs tend to be over-conservative when code reviewing, like a human software developer who tends to label a large number of code lines as suspect.
    \item H2 (lack of enough context): LLMs may lack extra context that otherwise would warrant correctness (e.g., correct variable initialization).
    \item H3 (dataset mislabeling): Some bugs might not have appeared at the time the dataset was originally built, but may have appeared later.  Therefore, some execution flow may not have been run, potentially causing a bug to go undetected.  The hypothesis is particularly important even for fine-tuned models because of the risk of overfitting to an outdated dataset.
\end{itemize}

For each of the above hypotheses, typical code snippets extracted from the JIT-Defects4J dataset are used to depict the scenario. 

\paragraph{H1 (human bias)}

According to this hypothesis, some errors may occur primarily because the LLM acts like a human being who is unaware of the execution configuration or the expected execution flow. Some concrete scenarios range from potential arithmetic errors to type-related errors.  In cases involving potential arithmetic errors, the LLM labeled as defect-inducing lines that perform operations such as dividing by a variable that could be zero, even though the code logic may guarantee that this situation never occurs. (Listing \ref{lst:arith_errors}).  In reflection/type-related errors, the LLM assumes that variable castings are potentially risky, even when they occur in controlled situations.  For example, in Listing \ref{lst:cast_error}, a potential casting error is reported, but at the current version of the ScriptObjectMirror class, it implements the Bindings interface\footnote{ScriptObjectMirror class documentation: \url{https://docs.oracle.com/javase/8/docs/jdk/api/nashorn/jdk/nashorn/api/scripting/ScriptObjectMirror.html}}.


\begin{listing}[htbp]
  \captionsetup{type=listing}
  \caption{Examples of \enquote{Potential arithmetic errors} error type}\label{lst:arith_errors}
\begin{lstlisting}[language=json]
{
  "lineNumber": 327,
  "codeContent": "sum += FastMath.pow(distA / distB, 2.0 / (fuzzyness - 1.0));",
  "reason": "Division by zero and NaN risk: if distB == 0 (point coincides with some cluster center), distA/distB is Infinity (or 0/0 -> NaN when ...",
  "project": "commons-math"
}

{
  "lineNumber": 159,
  "codeContent": "if (numUpdates % statsPrintFrequency == 0) {",
  "reason": "statsPrintFrequency can be 0 from configuration, causing ArithmeticException (modulo by zero). Previously this was a constant 10 and safe.",
  "project": "giraph"
}
\end{lstlisting}
\end{listing}

\begin{listing}[htbp]
  \captionsetup{type=listing}
  \caption{Examples of \enquote{Reflection/type-related errors} error type}\label{lst:cast_error}
\begin{lstlisting}[language=json]
{
"lineNumber": 321,
  "codeContent": "return (Bindings) bindings.get(NASHORN_GLOBAL);",
  "reason": "Unsafe cast: in Nashorn, 'nashorn.global' typically holds a ScriptObject/ScriptObjectMirror, not a javax.script.Bindings. This cast can trigger ClassCastException when accessed.",
  "project": "commons-scxml"
}
(...)
{
"lineNumber": 186,
  "codeContent": "return (Class<T>) getClassMethod.invoke(schema);",
  "reason": "Reflection is invoked on the wrong receiver. The method getClass(Schema) belongs to the model instance, but this calls it on the Schema object, causing IllegalArgumentException at runtime and preventing correct datum class detection.",
  "project": "parquet-mr"
}
\end{lstlisting}
\end{listing}

\paragraph{H2 (lack of enough context)}

Some errors may occur primarily due to a lack of context, such as false positives related to variable initialization and concerns about API versions.  For example, variables pointed as potentially null might have been correctly initialized before that point in the code.  Regarding concerns about API versions, constructors and method overloading may be available in older or more recent library versions, but the LLM does not have access to this information in the standard prompt format based on diffs.  This can lead, for example, to errors in the \enquote{Apparent compilation errors} and \enquote{Binary incompatibilities} categories.

\begin{listing}[htbp]
  \captionsetup{type=listing}
  \caption{Example of \enquote{Unchecked null access} error type}\label{lst:arith_errors_3}
\begin{lstlisting}[language=json]
{
  "lineNumber": 132,
  "codeContent": "Map parentPomProps = 
    PomModuleDescriptorBuilder.extractPomProperties(
      parentDescr.getExtraInfo());",
  "reason": "Dereferences parentDescr without verifying it is non-null. If the parent could not be resolved, this line will throw a NullPointerException.",
  "project": "ant-ivy"
}
\end{lstlisting}
\end{listing}

\begin{listing}[htbp]
  \captionsetup{type=listing}
  \caption{Example of \enquote{Apparent compilation errors} error type}\label{lst:arith_errors_3}
\begin{lstlisting}[language=json]
{
  "lineNumber": 436,
  "codeContent": "return new LZMACompressorInputStream(in, lzmaMemoryLimitKb);",
  "reason": "Unconditionally uses the new two-argument LZMA constructor after only checking LZMAUtils.isLZMACompressionAvailable(). That check verifies class presence but not constructor availability. With older org.tukaani.xz versions lacking LZMAInputStream(InputStream,int), this call can trigger NoSuchMethodError at runtime rather than a clean CompressorException.",
  "project": "commons-compress"
}
\end{lstlisting}
\end{listing}

\begin{listing}[htbp]
  \captionsetup{type=listing}
  \caption{Example of \enquote{Binary incompatibilities} error type}\label{lst:arith_errors_3}
\begin{lstlisting}[language=json]
{
  "lineNumber": 484,
  "codeContent": "return compressorStreamProvider.createCompressorInputStream(name, in, actualDecompressConcatenated);",
  "reason": "Calling the new 3-arg method on providers is binary-incompatible with existing third-party CompressorStreamProvider implementations compiled against the old 2-arg interface, causing AbstractMethodError/IncompatibleClassChangeError at runtime when discovered via service loading.",
  "project": "commons-compress"
}
\end{lstlisting}
\end{listing}

\paragraph{H3 (dataset mislabeling)}  The following errors may occur primarily due to poor dataset labeling because of error conditions that had not been observed at the time the dataset was labeled, such as, for example, a non-observable logic condition: the next listing shows an example of an execution flow that may not have been run, potentially causing a bug to go undetected.  In the example shown in Listing \ref{lst:logical_error}, clearly, if arrayObject is equal to null, the call to arrayObject.getClass()... will be tried, and a NullPointerException will be thrown.

\begin{listing}[htbp]
  \captionsetup{type=listing}
  \caption{Example of \enquote{Non-observable logic condition} error type}\label{lst:logical_error}
\begin{lstlisting}[language=json]
{
  "lineNumber": 121,
  "codeContent": "if (arrayObject != null && arrayObject instanceof Iterable || arrayObject.getClass().isArray()) {",
  "reason": "Operator precedence bug: when arrayObject is null, the right side arrayObject.getClass().isArray() is still evaluated, causing a NullPointerException. Should be: arrayObject != null && (arrayObject instanceof Iterable || arrayObject.getClass().isArray())",
  "project": "commons-scxml"
}
\end{lstlisting}
\end{listing}


\paragraph{False Negatives Analysis}

Regarding the second part of our investigation, GPT-5 generated a total of 844 false negatives, nearly half the number of false positives, supporting the intuition that GPT-5 favors a more conservative and bug-seeking approach (H1 in the false positives analysis).  

A preliminary analysis reveals common types of mistakes, as shown in Tables \ref{tab:fns-1} and \ref{tab:fns-2}.  Here, we opt to show the most intuitive ones instead of the most common ones.  In fact, there are a larger number of false negatives in code lines that contain cast operations (potentially unsafe), numeric computation, index-based control flow in loops, etc.  However, the goal of this analysis is to gather information on better prompt instructions. If we incorporate the full spectrum of false-negative categories into the prompt instructions, we risk generating yet more false positives. 

\begin{table}[t]
\centering
\scriptsize
\caption{Summary of False Negative Categories (Descriptions)}\label{tab:fns-1}
\begin{tabular}{|p{3.1cm}|p{6.3cm}|p{1.2cm}|}
\hline
\textbf{Category} & \textbf{ Description} & \textbf{Freq. (\%)} \\
\hline
I/O operations without validation &
Reading or writing data without checking return values or error conditions, risking partial reads or unnoticed failures. &
14 \\
\hline
Configuration or metadata merging &
Combining settings or metadata from multiple sources without proper validation or ordering, creating subtle inconsistencies. &
12 \\
\hline
Logging hides underlying issues &
Logging warnings or errors instead of handling exceptional situations masks real problems in the system. &
6 \\
\hline
\end{tabular}
\end{table}

\begin{table}[t]
\centering
\scriptsize
\caption{Representative Examples for Each False Negative Category}\label{tab:fns-2}
\begin{tabular}{|p{2.5cm}|p{8.4cm}|}
\hline
\textbf{Category} & \textbf{Representative Example (from dataset)} \\
\hline
I/O operations without validation &
\texttt{nbrBytesCopied += in.read(b, off + nbrBytesCopied, len - ...);} \\
\hline
Configuration or metadata merging &
\texttt{mergeConfigurations(parent.getModuleRevisionId(), parent.getConfigurations());} \\
\hline
Logging hiding underlying issues &
\texttt{LOG.warn("Ignoring statistics because created\_by...");} \\
\hline
\end{tabular}
\end{table}

\begin{tcolorbox}
\textit{[RQ3]: Despite their strong performance, the qualitative analysis shows that LLMs still underperform in three main situations: conservative human-like bias, where models over-flag lines as suspicious, leading to false positives such as theoretical arithmetic errors, overly cautious reflection/casting warnings, or misinterpreted logic conditions; insufficient contextual information, causing models to misjudge code that is actually safe or compatible due to missing details about variable initialization, API/version differences, or execution flow unavailable in diff-based prompts; and dataset noise or outdated labeling, which creates mismatches between ground truth and actual defect behavior.  Furthermore, the false negatives analysis reveals that LLMs may struggle in areas such as unvalidated I/O operations, fragile configuration/metadata merging, or logging patterns that conceal failures. Overall, these findings suggest that LLMs remain susceptible to context gaps and conservative reasoning patterns, highlighting opportunities for enhanced prompt design, context engineering, and refined dataset curation in the evaluation phase to mitigate such errors.  This qualitative taxonomy addresses a critical gap in the literature by offering the first systematic analysis of LLM reasoning failures in JIT-DL, providing actionable insights for prompt design, model evaluation, and dataset refinement. 
}
\end{tcolorbox}

\section{Conclusions}\label{sec:conclusions}

This work introduced CodeFlowLM, an incremental learning framework for Just-In-Time Software Defect Prediction (JIT-SDP) built upon pre-trained language models. Across within-project and cross-project experiments, CodeFlowLM — particularly when using CodeT5+ — consistently outperformed traditional incremental learners such as BORB in both predictive accuracy and stability, achieving superior G-Mean values in most evaluated projects. Our findings confirm, for the first time, that PLMs can be incrementally adapted under real JIT-SDP constraints—drift, verification latency, and imbalance evolution—without model restarts or costly hyperparameter searches.

We also extended our investigation to the Just-in-Time Defect Localization (JIT-DL) task, analyzing the performance and limitations of Large Language Models (LLMs) in this context. Among the evaluated models, GPT-5 achieved competitive results compared to attention-based approaches such as JIT-Smart\textsubscript{ATTN}, particularly in the Recall@20\% and Effort@20\% metrics, while exhibiting greater stability between runs. Nevertheless, attention-based models still outperformed LLMs in ranking-oriented measures such as Top-5 and IFA, suggesting that although LLMs show strong potential for defect localization, further progress is needed to enhance their fine-grained line prioritization capabilities.

Critically, our qualitative investigation fills a key gap in the literature: understanding why LLMs fail.  Our analysis revealed that false positives occur at a higher frequency than false negatives.  The false positives produced by LLMs are often caused by (1) human-like conservative bias, leading to over-identification of suspicious lines; (2) insufficient contextual information due to limited diff-based prompts; and (3) dataset labeling noise, where some “clean” commits in JIT-Defects4J should, in fact, be labeled as defect inducing. These findings highlight both the potential and the current limitations of generative models in reasoning about software changes.

Future work will explore drift-adaptive thresholds, multimodal fusion for joint prediction/localization, and an automated audit of JIT-Defects4J labels. Together, these steps move us toward unified, robust, and trustworthy JIT defect analysis pipelines.


%
%


%
%

\section*{Declarations}
\raggedright

\textbf{Funding}: Not applicable.

\textbf{Ethical Approva}l Not applicable. 

\textbf{Informed Consent} Not applicable. 

\textbf{Author Contributions}: Monique Louise Monteiro formulated research questions and hypotheses and conducted the experiments.  George G. Cabral and Adriano L.I. Oliveira contributed to the formulation of research questions and hypotheses, as well as the discussion of results, and reviewed the manuscript.

\textbf{Data Availability Statement}: The reproduction package, including source code and data used in the experiments, is available at \url{https://github.com/monilouise/codeflowlm_jitdl}.

\textbf{Conflicts of Interest} The authors declared that they have no conflict of interest.

\textbf{Clinical Trial Number} Not applicable.

\bibliographystyle{spbasic}      
\bibliography{biblio}


%
%

\end{document}